\documentclass[prl,twocolumn,showpacs,floats,floatfix,nofootinbib]{revtex4}

\usepackage{natbib}
\usepackage{amsbsy}
\usepackage{amssymb}
\usepackage{amsmath}

\usepackage{graphicx}

\def\n{{\mathrm{n}}}

\def\n{{\rm n}}
\def\p{{\rm p}}

\def\c{{\rm c}}

\def\n{{\rm n}}
\def\p{{\rm p}}

\def\c{{\rm c}}

\def\be{\begin{equation}}
\def\ee{\end{equation}}
\def\beq{\begin{equation}}
\def\eeq{\end{equation}}
\def\bea{\begin{eqnarray}}
\def\eea{\end{eqnarray}}

\def\bear{\begin{eqnarray}}
\def\eear{\end{eqnarray}}

\begin{document}

\title{Pulsar  glitches: The crust is not enough}

\author{N. Andersson$^1$, K. Glampedakis$^{2,3}$, W.C.G. Ho$^1$ and  C.M. Espinoza$^4$} 

\affiliation{$^1$School of Mathematics, University of Southampton, Southampton SO17 1BJ, UK}
\affiliation{$^2$Departamento de F\'isica, Universidad de Murcia, Murcia E-30100, Spain \\
$^3$Theoretical Astrophysics, University of T\"ubingen, T\"ubingen D-72076, Germany,}
\affiliation{$^4$Jodrell Bank Centre for Astrophysics, School of Physics and Astronomy,
 University of Manchester, Manchester M13 9PL, UK}


\begin{abstract}

Pulsar glitches are traditionally viewed as a manifestation of vortex dynamics associated with a neutron
superfluid reservoir confined to the inner crust of the star. In this Letter we show that the 
non-dissipative entrainment coupling between the neutron superfluid and the nuclear lattice leads to a less mobile 
crust superfluid, effectively reducing the moment of inertia associated with the angular momentum reservoir. Combining the latest observational data for prolific glitching pulsars with theoretical results for the crust entrainment 
we find that the required superfluid reservoir exceeds that available in the crust. This challenges our understanding of the glitch phenomenon, and we discuss possible resolutions to the problem.
\end{abstract} 

\maketitle


{\em Context}.-- Mature neutron stars tend to be extremely stable rotators, in some cases rivalling the best terrestrial 
atomic clocks. However, in their adolescence they may behave in a less ordered fashion. In particular, many young neutron 
stars exhibit (more or less) regular glitches, where the observed spin rate suddenly increases (for a recent collection of
glitch data, see~\cite{espinoza11}). These spin-up events tend to be followed by a slow relaxation 
towards the original spin-down rate. In some cases (notably the Crab pulsar) the glitch leads to a permanent change in 
the spin-down rate, but in most cases the system resumes spinning down as before the event.

The archetypal glitching neutron star is the Vela pulsar, which has (since the first observed event in 1969) exhibited a reqular sequence of similar size glitches. The consensus view is that these events are a manifestation of the presence of a superfluid component in the star's interior~\cite{baym69}. This idea was first put forward by Anderson \& Itoh \cite{ai75} who envisaged a glitch as a tug-of-war between the
tendency of the neutron superfluid to match the spindown rate of the rest of the star
by expelling vortices and the impediment experienced by the moving vortices due to ``pinning'' to crust nuclei. Strong vortex pinning prevents the
neutron superfluid from spinning down, creating a spin lag with respect to the rest of the star (which is spun down electromagnetically). 
This situation cannot persist forever. The increasing spin lag leads to a build up in the Magnus force exerted on the vortices. 
Above some threshold pinning can no longer be sustained, the vortices break free and the excess angular momentum is transferred
to the crust. This leads to the observed spin-up. 

Several decades have passed since the two-component model was first suggested, yet there has been surprisingly little progress on making the model quantitative. Most thinking has gone into the microphysics, especially concerning the  interaction between  neutron vortices and  crust  nuclei \cite{pizza}, a key ingredient in the scenario. Meanwhile,  the detailed mechanism responsible for triggering 
glitches in the first place remains unknown \cite{gaprl}. Moreover, we do not actually know the location of the superfluid reservoir associated with these events. It is generally assumed that the vortices are pinned in the crust and hence the angular momentum available is that associated with the crust superfluid. This notion is supported by data for frequent glitchers (Vela being the prime example). The analysis by Link et al~\cite{link99} suggests that
glitches represent a self-regulated process that involves a superfluid reservoir with moment of inertia 
$I_{\rm n}/I \sim 1\%$ where $I$ and $I_{\rm n}$ are, respectively, the moments of inertia of the entire star and the
superfluid component. The similarity of the inferred $I_{\rm n}$ to the theoretically estimated moment of inertia of the crust (which is dominated by the free neutrons in the inner crust) for realistic equations of state \cite{rp94}   
supports the notion that glitches involve only the crust region.

We argue that this logic breaks down when one accounts for the non-dissipative entrainment coupling between the neutron superfluid and the crust lattice, an effect which can be expressed in terms of an effective neutron mass, $m_\n^*$. Recent work indicates that this effective mass may be significantly larger than the bare neutron mass, $m_\n$ \cite{chamel05, chamel12}. This implies a decreased mobility of the superfluid with respect to the lattice and the need for a larger angular momentum reservoir for glitches. 
Combining the latest data for glitching systems~\cite{espinoza11} with a general relativistic multifluid model that includes entrainment, we find that the required superfluid moment of inertia 
is above the capacity of the crust superfluid. This suggests that
some fraction of the core superfluid must participate in glitches, which raises a number of interesting questions.  
 
{\em Phenomenology and observations}.--
The discussion of vortex mediated glitches is usually based on a ``body'' averaged model with two components. The first represents the charged component (including the elastic crust) which is spun down electromagnetically. Labelling this component by an index p,  we have
\be
I_\p \dot \Omega_\p  = -a \Omega_\p^3 -\mathcal{N}_\mathrm{pin} -\mathcal{N}_\mathrm{MF}
\label{peq}\ee
where the first term on the right-hand side represents the standard torque due to a magnetic dipole (the coefficient $a$ depends on the moment of inertia, the magnetic field strength and its orientation; we assume that these parameters  do not evolve with time).
We also have a superfluid component, with index n, which evolves according to
\be
I_\n \dot \Omega_\n  = \mathcal{N}_\mathrm{pin} +\mathcal{N}_\mathrm{MF}
\label{neq}\ee
On the right-hand sides of these equations we have added terms representing torques associated with
vortex pinning ($\mathcal N_\mathrm{pin}$) and dissipative mutual friction ($\mathcal N_\mathrm{MF}$) associated with scattering off of the vortices in the superfluid. We will not need explicit forms for these in the following.

Glitches can be understood as a two-stage process. In the first phase the superfluid vortices are pinned. This means that $\mathcal N_\mathrm{pin}$ is exactly such that $\dot\Omega_\n=0$. That is, the pinning force counteracts the friction which tries to bring the fluids into co-rotation. The upshot is that the crust evolves according to 
\be
I_\p \dot \Omega_\p  = -a \Omega_\p^3 \ \longrightarrow \ 
{1\over \Omega_\p^2} - {1\over \Omega_0^2} = {2 a \over I_\p}  (t-t_0)
\ee
Assuming that a system starts out at co-rotation (with spin $\Omega_0$ at time $t_0$), we can estimate how the spin-lag, $\Delta \Omega = \Omega_\n - \Omega_\p$, between the two components evolves with time.
As long as the spin-lag is small we have $\Delta \Omega/\Omega_\p \approx { t_\mathrm{glitch}/ 2 \tau_c}$ 
where $t_\mathrm{glitch}$ is the interglitch time and $\tau_c = - { \Omega_\p / 2 \dot \Omega_\p}$ is the characteristic age of the pulsar.

At some point, this lag reaches a critical level where the vortices unpin. The two components then relax to co-rotation on the mutual friction timescale (which may be as fast as a few hundred rotations of the system \cite{alpar}).
This transfers angular momentum from the superfluid reservoir to the crust, leading to the observed glitch. Assuming that angular momentum is conserved in the process (such that the entire spin-lag $\Delta \Omega$ drives the observed glitch jump $\Delta \Omega_\p$) we have
\be
I_\p  \Delta \Omega_\p = I_\n \Delta \Omega \quad  \longrightarrow \quad
{\Delta \Omega_\p  \over \Omega_\p } \approx {I_\n \over I} {t_\mathrm{glitch}\over 2 \tau_c} 
\label{eq4}\ee
where $I=I_\n+I_\p $ is the total moment of inertia (we have assumed a small superfluid reservoir,
i.e. $I \approx I_\p$).

\begin{table}[h]
\begin{tabular}{|l|c|c|c|}
\hline
PSR &  $\tau_c$ (kyr)	& $\mathcal A$ ($\times 10^{-9}$/d) & $ I_\n/I $ (\%)  \\
\hline
J0537-6910		& 4.93 & 2.40 & 0.9 \\
B0833-45 (Vela)  & 11.3 & 1.91 &  1.6 \\
\hline
J0631+1036 & 43.6 & 0.48 & 1.5 \\
B1338-62 & 12.1 & 1.31 & 1.2  \\
B1737-30  & 20.6 & 0.79 & 1.2 \\
B1757-24  &  15.5 & 1.35 &  1.5 \\
B1758-23  & 58.4 & 0.24 & 1.0 \\
B1800-21  & 15.8 & 1.57 & 1.8 \\ 
B1823-13  & 21.5 & 0.78 &  1.2  \\
B1930+22  & 38.8 & 0.95 & 2.7 \\
B2229+6114 & 10.5 & 0.63 & 0.5 \\
\hline
\end{tabular}
\caption{Inferred superfluid moment of inertia fraction for  glitching pulsar which have exhibited at least two (large) events of similar magnitude. The data is taken from \cite{espinoza11} (updated to included a few more recent events \cite{update}), c.f., Figures~\ref{fig1} and \ref{fig2}. We give each pulsars name, the characteristic age, $\tau_\c$, the averaged rate of spin-reversal due to glitches,  $\mathcal A$, and the moment of inertia ratio $I_\n/I$ obtained from \eqref{eq6}. }
\label{tab1}
\end{table}

Let us compare this model to observations. To do this, we assume that we see a number of glitches in a given system during an observation campaign lasting $t_\mathrm{obs}$.
Then we can work out the accumulated change in the observed spin due to glitches, and relate the result to the simple two-component model. 
From \eqref{eq4} we then have
\be
{I_\n /I} \approx 2 \tau_c \mathcal{A} \quad \mbox{where} \quad \mathcal{A} = {1 \over  t_\mathrm{obs}} \left( {\sum_i \Delta \Omega_\p^i/ \Omega_\p}  \right)
\label{eq6}\ee
For systems that have exhibited at least two glitches of similar magnitude \cite{espinoza11} we can estimate the average reversal in spindown due to (large) glitches per day  of observation, $\mathcal A$.  This leads to the inferred moment of inertia fractions listed in Table~\ref{tab1}. For some systems, like the Vela pulsar and the X-ray pulsar J0537-6910, the estimate should be quite reliable given the number of glitches exhibited and their regularity. In other cases, the data is less impressive, as is clear from Figure~\ref{fig2}. Nevertheless, the message seems clear: Glitches require the superfluid component to be associated with at least 1-1.5\% of the star's moment of inertia. This agrees with the conclusions of \cite{link99}. In addition, the data seems consistent with the idea of an angular momentum reservoir that is completely exhausted in each event. If this is not the case then it is difficult to explain why the recurring glitches have such similar magnitude. 

\begin{figure}[h]
\includegraphics[width=8cm,clip]{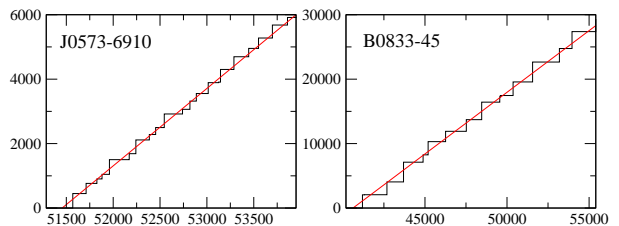}
\caption{The accumulated $\sum_i \Delta \Omega_\p^i/\Omega_\p$ ($\times 10^{-9}$) as a function of Modified Julian date for the X-ray pulsar J0537-6910 and the Vela pulsar (B1833-45). The fits that lead to the slopes $(\mathcal A$) listed in Table~\ref{tab1} are shown as straight lines. }
\label{fig1}
\end{figure}

\begin{figure}[h]
\includegraphics[width=8cm,clip]{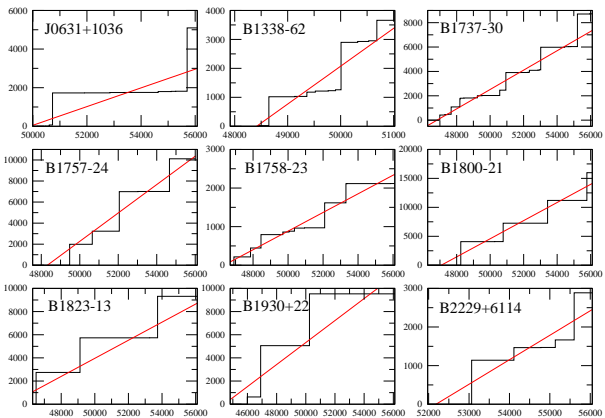}
\caption{Same as Figure~\ref{fig1},  for pulsars with a smaller number of large glitch events.}
\label{fig2}
\end{figure}

{\em The role of entrainment}.-- Let us now ask what the influence of a ``heavy'' superfluid may be. That is, let us account for the entrainment coupling. At the level of the averaged two-component model, the entrainment can be expressed in terms of a 
coefficient $\varepsilon_\n$. The two equations of motion then take the form \cite{living}
\be
\left( I_\p - \varepsilon_\n I_\n\right) \dot \Omega_\p + \varepsilon_\n I_\n \dot \Omega_\n = -a \Omega_\p^3  -\mathcal{N}_\mathrm{pin} -\mathcal{N}_\mathrm{MF}
\label{Jpdot}\ee
and
\be
\left(1 - \varepsilon_\n\right) I_\n \dot\Omega_\n + \varepsilon_\n I_\n \dot \Omega_\p =  \mathcal{N}_\mathrm{pin} + \mathcal{N}_\mathrm{MF}
\label{Jndot}
\ee
Combining these (noting that  the right-hand side of (\ref{Jndot}) vanishes for perfect pinning),
we see that the crust now spins down according to
\be
\tilde I \dot \Omega_\p = - a \Omega_\p^3 \quad \mbox{where} \quad 
\tilde I = I_\p - {\varepsilon_\n \over 1 - \varepsilon_\n } I_\n 
\ee
Expressing the entrainment in terms of the (average) effective neutron mass, we have
\be
\varepsilon_\n = 1 - { \left<  m_\n^* \right> \over m_\n}  \ \longrightarrow \ 
\tilde I = I -{m_\n \over  \left<  m_\n^* \right>}  I_\n 
\label{epsdef}\ee
The interpretation of this result is quite simple. The entrainment encodes the mobility of the superfluid neutrons relative to the other component. If the effective mass is large, then the two components are effectively locked. Hence, the system spins down as one body ($\tilde I \to I$)  in the limit where $\langle m_\n^* \rangle  \gg m_\n$. Basically, the entrainment  lowers the ``effective'' moment of inertia associated with the superfluid.

In terms of the observed spin-down the entrainment only has the effect of altering the inferred magnetic field. We can still introduce the characteristic age (obtained from observables) to get the accumulated spin-down of the crust. 
The main difference now is that $\Omega_\n$ also changes (even when vortices are pinned). From \eqref{Jndot} we have 
\be
\dot\Omega_\n = - {\varepsilon_\n \over 1 - \varepsilon_\n } \dot \Omega_\p = \left( 1 - {m_\n \over \langle m_\n^* \rangle} \right) \dot \Omega_\p
\ee
This has repercussions for the estimated glitch jumps because the spin-lag between the two components takes a longer time to develop
if the effective neutron mass is large. Working out the accumulated spin-lag and assuming angular momentum conservation during the glitch, we have (again assuming $I \approx I_\p$)
\be
{\Delta \Omega_\p \over \Omega_\p} \approx {m_\n \over \langle m_\n^* \rangle }  \left({I_\n \over I}\right) { t_\mathrm{glitch} \over 2 \tau_c}
\ee

The observations then provide us with the constraint
\be
{I_\n \over I} \approx  2 \tau_c  \mathcal{A}  
 {\langle m_\n^* \rangle \over m_\n}  
\label{SFratio}
\ee
In other words, if the (average) effective neutron mass is large, then the constraint inferred from glitch observations will be more severe than previously assumed (e.g., in \cite{link99}). This argument may not be new \cite{carter}, but the effect has not previously been quantified.


{\em Moments of inertia}.--
In order to quantify the constraint set by the glitch data, we need a relativistic model
for the involved moments of inertia \cite{nagc01}. The need for such model is emphasised by the recent results of Chamel \cite{chamel05,chamel12},
which suggest that the effective mass for the superfluid neutrons that permeate the inner crust may, indeed, be quite large. 
The phenomenological (body averaged) entrainment model from the previous section illustrates how a large effective mass  affects the analysis, but we need to connect this argument with a detailed neutron star model, incorporating realistic crust and core physics, as well as reasonable superfluid parameters. We build on the relativistic superfluid formalism developed in \cite{nagc01}, and use the same assumptions regarding the supranuclear equation of state and the singlet pairing gap for the neutrons as in \cite{magcool}. The effective neutron mass in the crust is estimated using the phenomenological fit to the entrainment from \cite{magnetar}. We consider two models, based on \cite{chamel05} and \cite{chamel12}, respectively. The main difference is that the effective mass peaks at lower densities in the latter case. This serves to weaken the effect we are discussing slightly. 

Defining the moment of inertia through $J=I\Omega$, we get the total moment of inertia from \cite{rp94} 
\be
I \approx \left( 1 - {2I \over R^3} \right)  I_0
\ee
where $R$ is the radius of the star, and
\be
I_0 = {8 \pi \over 3} \int_0^R  r^4 e^{(\lambda - \nu)/2}
         \left( p+\rho \right) dr 
\ee
where $p$ and $\rho$ are the pressure and the energy density, respectively, and $\lambda$ and $\nu$ determine the spacetime metric.
 In the case of the superfluid crust neutrons we ignore the effect of the rotational frame-dragging (this should be, at most, a 20\% correction). Making contact with the phenomenological model discussed previously, we express the superfluid moment of inertia as (see \cite{nagc01})
\be
J_\n = I_\n \Omega_\n + \varepsilon_\n I_\n  \left(\Omega_{\p} - \Omega_{\n}\right) 
\label{Jn}
\ee
Here 
\be
I_\n \approx {8 \pi \over 3}  \int_{R_c}^R r^4 e^{(\lambda - \nu)/2}
        n_\n \mu_\n dr 
\ee
where $n_\n$ is the number density of the free neutrons and $\mu_\n$ is the corresponding chemical potential, and
\be
 \varepsilon_\n = {1 \over I_\n } {8 \pi \over 3}  \int_{R_c}^R r^4 e^{(\lambda - \nu)/2}
         n_\n \left(m_\n -  m_\n^*\right) dr
\ee
In these integrals we only account for the crust superfluid ($R_c$ represents the crust-core interface), but it is obviously straightforward to introduce a core component as well (as long as we keep in mind that the entrainment is rather different in the core).
In the inner crust we can safely assume that the free neutrons  are non-relativistic, which means that $\mu_\n \approx m_\n$. It is then straightforward to work out the effect of the entrainment (in terms of $\varepsilon_\n$, or the averaged ratio $\langle m_\n^* \rangle /m_\n $ via \eqref{epsdef}) and check to what extent a given neutron star model satisfies the constraints set by the observations. Let us focus on the results obtained for the  equations of state for  core and crust used in \cite{magcool}. For this model we find that $I_\n/I$ ranges from around 6\% for a $1.2M_\odot$ star to 4\% for a 1.4 $M_\odot$ star and less than 2\% for a 1.8$M_\odot$ star. This means that, as long as we do not worry about the entrainment, these models easily satisfy the constraints set by the observed systems; The angular momentum reservoir exceeds the requirements. Testing a few different core equations of state we find that this conclusion holds in general. It is worth noting that the $I_\n/I$ ratio is  significantly smaller for the models considered in \cite{rp94,link99}. Turning to the entrainment effect, we find that the two models (representing the results from \cite{chamel05} and \cite{chamel12}) lead to $\langle m_\n^*\rangle /m_\n$  in the range $4-6$ (more precise estimates are not justified given that we are using a phenomenological fit to the small number of data points given in \cite{chamel05,chamel12}). At this point the standard glitch logic is in trouble. Unless the observed systems are all low mass neutron stars, the crust is not enough; The associated superfluid does not have sufficient moment of inertia to explain the observations.

{\em Discussion.}-- Our analysis casts  doubt on the standard model of pulsar glitches being
driven by a superfluid reservoir confined to the inner crust of the star. This seems inevitable provided the effective mass associated with entrainment is indeed as large as suggested by recent work~\cite{chamel05,chamel12}. Basically, the 
entrainment  lowers the effective superfluid to total moment of inertia by a factor $\sim 4-6$ below that assumed in previous work~\cite{link99}. This brings the theory into conflict with the data for pulsars with regular large glitches, as shown in Table~\ref{tab1}. In order to be ``consistent'' we need stellar models with 
$I_\n/I \gtrsim 6-10 \% $, but this is only borderline achievable for proposed realistic equations of state provided that these neutron stars  all have low mass.

There are (at least) three possible interpretations of this analysis. The first possibility is, perhaps, the least attractive. The neutron star core is expected to 
contain superfluid neutrons in abundance. In the case of the singlet pairing gap we have considered (see \cite{magcool}) the total  moment of inertia fraction would be at least a factor of two (up to an order of magnitude for massive stars) larger than required (and one should probably  add the triplet pairing region to this).  This is  
where a new problem arises. The remarkable regularity of the glitches in systems like Vela and PSR J0537-6910 
suggests a, more or less, completely recycled reservoir of pinned superfluid ---  the very  reason why the previous
moment of inertia constraints~\cite{link99} were taken as evidence in favour of crust-only based glitches. 
However, if the core superfluid takes part, then why are the glitches not larger (at least occasionally)? Could it be that the unknown glitch trigger mechanism is associated with some characteristic lengthscale that leads to only a fraction of the total reservoir 
being replenished in each event? To make progress on these questions, we need more detailed work on plausible glitch triggers (possibly associated with the onset of superfluid turbulence \cite{gaprl,link12}).

The second possible explanation is that the core superfluid is involved in the glitch, but that the combined superfluid reservoir is just large enough to explain the observations. If this ``fine-tuning'' resolves the problem, then a more detailed calculation would be able to constrain the singlet pairing gap for neutrons. This would be a very interesting complement to the recent constraints on the core superfluids obtained from the cooling rate of the young neutron star in the Cassiopeia A supernova remnant \cite{casa,casb}.

The final possibility is, conceptually, the simplest. The discrepancy between the model and the observations may be due to a lack of ``precision'' in the theory. This is obviously testable. However, a more precise analysis of the problem requires a consistent crust-core equation of state along with the effective neutron mass for the superfluid component. So far the only results in this direction are those in \cite{chamel05,chamel12}. Work aimed at confirming (or refuting) those results should be strongly encouraged. 

There is, however, a problem also with this explanation. Unless the superfluid is confined to the crust, one would have to explain why the crust component decouples from the core during the glitch event. This would be particularly vexing if the singlet pairing gap is such that the neutron superfluid reaches far into the core (which is the case in the model we have considered), leading to interesting questions concerning the physics associated with the transition from crust to core. A central issue concerns the nature of superfluid vortices extending across this interface. The standard picture is that the vortices are magnetized in the core (due to entrainment and the presence of superconducting protons) \cite{alpar} but not in the crust.  This hints at a more complicated behaviour at the interface than is usually assumed. Perhaps there is an argument for why the crust component  decouples. This problem needs further consideration.

Our analysis points to a number of interesting directions for future work. There is an obvious need for 
more precise theoretical limits on $I_\n/I$, both from theory and observation. On the theory side, the effective neutron mass in the crust should be a key ingredient in any model. From observations, continuous monitoring of glitching pulsars will improve the constraints on   
the superfluid reservoir. We also need to develop a better understanding of 
the superfluid/vortex physics at the crust-core interface. Progress along these lines is required if we want to take our understanding of glitches
to the next level, and use precision observations from the next generation of observational facilities to probe superfluid neutron star physics.

\acknowledgements
NA and WCGH are supported by STFC in the UK. KG is supported by the Ram\'{o}n y Cajal Programme in Spain. 
CME's pulsar research at JBCA is supported by a STFC Rolling Grant in the UK.


\end{document}